# A Curved Monopole Antenna for HF Radar with Enhanced Gain and Bandwidth

Masoud Salmani Arani
Department of Electrical and Computer Engineering
Memorial University of Newfoundland
St. John's, Canada
msalmaniaran@mun.ca

Reza Shahidi
Department of Electrical and Computer Engineering
Memorial University of Newfoundland
St. John's, Canada
rshahidi@mun.ca

Lihong Zhang
Department of Electrical and Computer Engineering
Memorial University of Newfoundland
St. John's, Canada
lzhang@mun.ca

*Abstract*— **This paper presents the design and simulation of a new curved monopole antenna optimized for skywave HF radar applications, with a systematic investigation of the effects of curvature and fixed-section length on antenna performance. The proposed design achieves improved impedance matching, broader bandwidth, and enhanced realized gain compared to a conventional quarter-wavelength monopole at 15 MHz. Parametric analysis shows that fully bending the monopole degrades performance, whereas introducing a straight section and carefully optimizing the curvature enables a 18.5% gain increase and a 400 kHz bandwidth expansion. The single-element design is further extended to a 12-element linear array with 0.45λ spacing (where λ is the wavelength), demonstrating stable embedded-element behavior and improved low-to-moderate elevation gain for skywave over-the-horizon radar operation. At $\theta = 30°$, the proposed array achieves 14.04 dBi compared to 13.11 dBi for the reference array, corresponding to 24% gain enhancement, which is significant in high-power HF radar systems. These results confirm that the proposed curved monopole antenna provides a compact, broadband, and scalable solution for next-generation HF radar arrays.**

## I. Introduction

High-frequency (HF) radar systems, typically operating in the 3–30 MHz frequency band, have served as a critical technology for long-range sensing and surveillance since the mid-20th century. Unlike conventional radars that are limited by the Earth's horizon, HF radars utilize unique propagation modes to achieve Over-the-Horizon (OTH) capabilities [1]. These systems primarily function through two methods: skywave propagation, which uses the ionosphere as a mirror to monitor regions thousands of kilometers away, and surface wave propagation, which leverages diffraction to follow the Earth's curvature for maritime surveillance up to approximately 400 km [2].

The utility of these propagation characteristics extends their applications beyond defense into environmental science, particularly for ice-penetrating radar in polar regions. Because the propagation of radio waves through cold ice is effectively frequency-independent in the HF and VHF bands, these radars are essential for monitoring ice-shelf stability and subsurface topography [3]. Modern applications often integrate these radars with autonomous rovers or utilize Synthetic Aperture Radar (SAR) techniques to study temporal changes in basal ice structures, requiring antennas that can maintain phase coherence and adequate gain [4].

Despite these advantages, the primary challenge in HF radar design remains the large physical size of the antennas, which is a direct consequence of the long wavelengths at these frequencies. For mobile or space-constrained applications, engineers must often resort to electrically small antennas, which frequently suffer from high capacitive reactance, low efficiency, and narrow bandwidths [5], [6]. To mitigate these issues, antenna arrays are utilized to provide necessary directivity and gain, though they often introduce further complexity regarding mechanical stability and management of grating lobes [7].

Current research has explored various configurations to optimize performance, including top-loaded umbrella monopoles, log-periodic dipole arrays (LPDAs), and cavity-backed structures. While horizontally-polarized LPDAs can offer wide bandwidth, they often require significant height and transportable towers for effective gain [8], [9]. Conversely, vertical monopole arrays are favored for surface wave applications but require careful impedance matching and tuning to maintain efficiency when reduced in size. There is a continuous demand for novel geometries that can deliver high gain and broad bandwidth without the prohibitive footprint of traditional HF structures [10].

This paper presents the development of a curved monopole antenna array designed to address these fundamental trade-offs by achieving improved bandwidth and gain while reducing the physical height of the monopole, offering a more efficient solution to modern HF radar applications.

The remainder of this paper is organized as follows: Section II describes the antenna design methodology. Section III presents the parameter analysis. Section IV discusses the simulation results and performance comparison. Section V details the array configuration and its radiation characteristics; and Section VI concludes the paper.

## II. Antenna Design Methodology

The proposed monopole antenna shown in Fig. 1 is composed of two geometrically distinct sections: a lower straight segment of length $L_{straight}$, referred to as the fixed section, and an upper curved segment that modifies the overall current path. This separation demands a systematic investigation of the impact of curvature while maintaining a stable and well-defined feed and reference structure. The fixed section establishes the electrical connection to the ground plane and sets the vertical reference height, while the curved section introduces geometric variation without increasing the antenna's total electrical length.

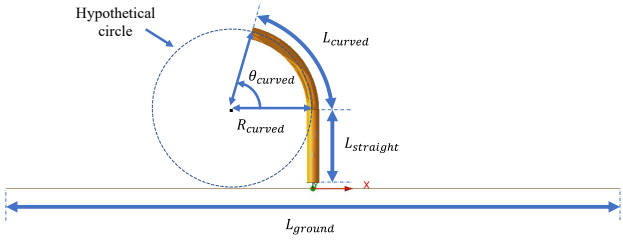

Fig. 1. Proposed curved monopole antenna.

The curved portion of the monopole is modeled using a three-dimensional tubular arc, defined by a major radius $R_{Curved}$. This parameter directly controls the curvature of the antenna and provides a continuous transition from a straight monopole to increasingly curved geometries. The curvature $\kappa$ is therefore defined as

$$\kappa = \frac{1}{R_{curved}} \quad (1)$$

such that a larger value of $R_{curved}$ correspond to a gentler curvature, and the limiting case of $R_{curved} \to \infty$ represents a conventional straight monopole.

To ensure fair comparison between different geometrical configurations, the total electrical length of the antenna is kept constant across all designs. This constraint is enforced by maintaining the sum of the fixed straight section length ($L_{straight}$) and the arc length of the curved section ($L_{curved}$) equal to the length of a reference straight monopole, $L_{ref}$. This condition is expressed as

$$L_{ref} = L_{straight} + L_{curved} \quad (2)$$

where the curved section length is given by

$$L_{curved} = R_{curved} \times \theta_{curved} \quad (3)$$

where $\theta_{curved}$ denotes the subtended angle of the curved segment. For each value of $R_{curved}$, the corresponding $\theta_{curved}$ is adjusted to preserve a constant total monopole length, ensuring that observed performance variations are attributed primarily to curvature effects rather than changes in electrical length.

This formulation treats curvature $\kappa$ and straight section length $L_{straight}$ as independent design parameters, providing a clear geometric basis for assessing curvature-induced changes in the current path and reactive behavior. The effects of $L_{straight}$ and curved length $L_{curved}$ on the performance of the antenna are investigated in the following section.

## III. PARAMETER ANSLYSIS

To evaluate the impact of the monopole geometry on antenna performance, two main parameters were investigated: the curvature of the curved section ($\kappa$) and the length of the straight section ($L_{straight}$). In the first step, $L_{straight}$ was fixed at 200 cm while the curvature of the monopole was varied. As shown in Fig. 2, increasing the curvature up to $\kappa = 0.5$ leads to an improvement in impedance matching, characterized by lower return loss and increased bandwidth. However, further increases in curvature beyond this value result in performance degradation, with worsening return loss and reduced bandwidth. These results indicate the existence of an optimal curvature range, beyond which excessive bending introduces unfavorable reactive effects that negatively impact antenna performance.

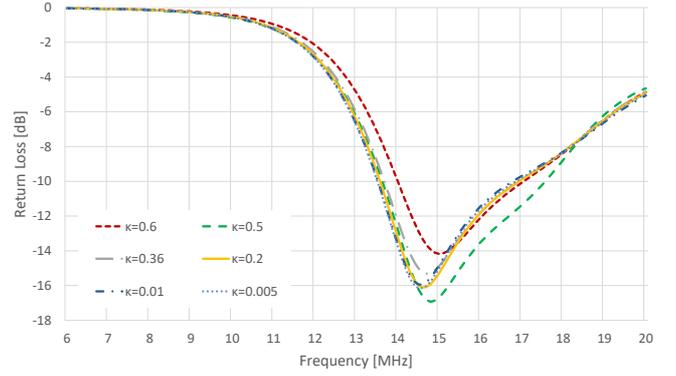

Fig. 2. Return loss vs frequency for different curvature values.

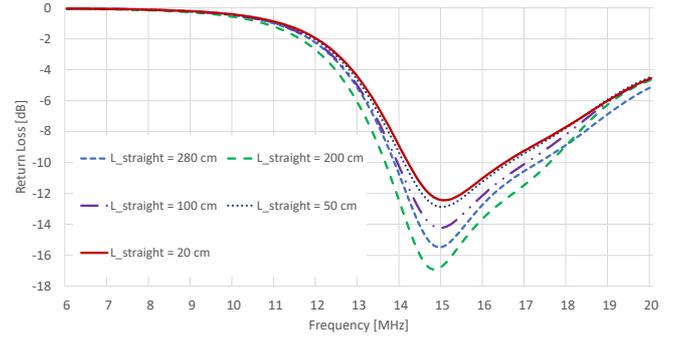

Fig. 3. Return loss vs frequency for different $L_{straight}$ values.

Subsequently, with the curvature fixed at its optimized value, the impact of the straight section length was examined. Fig. 3 shows the simulated return loss and bandwidth for different values of $L_{straight}$. Increasing $L_{straight}$ initially improves impedance matching and expands the operational bandwidth, indicating that a sufficiently long straight section stabilizes the input impedance of the antenna. This improvement persists up to $L_{straight} = 200cm$, beyond which further increases in length result in degraded return loss. These results confirm the existence of an optimal balance between the straight and curved portions of the monopole, where the straight section ensures impedance stability while the curved section provides controlled electrical length extension for broadband performance.

Overall, the parametric analysis highlights that neither an entirely curved monopole nor an excessively-long straight section is ideal. Instead, an intermediate configuration, combining a properly-sized straight segment with a carefully-optimized curved section, achieves the best trade-off between bandwidth, return loss, and antenna compactness. These findings provide clear design guidelines for tailoring the geometry of curved monopole antennas in HF radar applications.

## IV. RESULTS AND DISCUSSION

The optimized design of the proposed curved monopole was evaluated in comparison with a conventional straight monopole. All simulations were performed assuming a finite perfectly conducting ground plane of size 15 m × 15 m. As a reference, the length of a conventional monopole is approximately determined by the quarter-wavelength relation:

$$L_{ref} = \frac{\lambda}{4} = \frac{c}{4f_c} \quad (4)$$

where $c$ is the speed of light and $f_c$ is the center operating frequency. For the target HF operating band with $f_c = 15 MHz$, this yields a reference monopole length of approximately $L_{ref} = 467 cm$.

The optimized curved monopole, with a straight section length of $L_{straight} = 200 cm$ and a curved section characterized by a curvature radius $R_{curved} = 200 cm$ (or $\kappa = 0.5 m^{-1}$) and curved length of $L_{curved} = 267 cm$ and $\theta_{curved} = 1.33\ rad$ demonstrates significant performance improvements over the reference design. The maximum realized gain increased from 3.21 dBi for the conventional monopole to 3.95 dBi for our design as shown in Fig. 4, representing a 18.5 % enhancement. This gain improvement is attributed to the more favorable current distribution along the curved path, which leads to more efficient radiation and a slightly more directive pattern. In addition, Fig. 5 compares the radiation pattern of the proposed and reference antennas. It is noted that, due to the finite size of the ground plane, the radiation characteristics deviate from the ideal half-space behavior predicted by image theory for an infinite ground plane. Therefore, the reported gain values correspond to the practical finite-ground configuration rather than the idealized case.

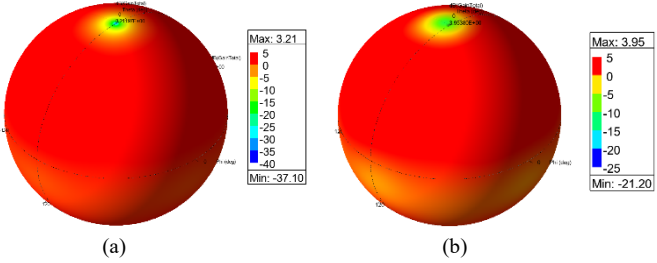

Fig. 4. 3D polar plot of the realized gain of the a) reference monopole antenna, and b) proposed curved monopole antenna.

In addition to gain enhancement, the optimized design exhibits improved impedance matching as shown in Fig. 6. The return loss shows a slight reduction, indicating better input matching across the operating band. As a result, the usable bandwidth is extended by approximately 400 kHz around the 15 MHz center frequency, demonstrating that the curvature and the inclusion of a straight section effectively

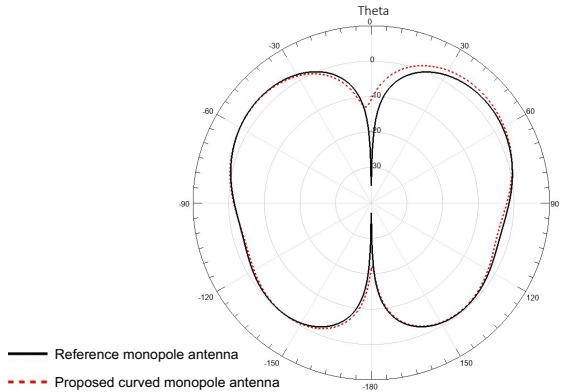

Fig. 5. Radiation pattern plot comparison in the $\varphi = 0$ plane.

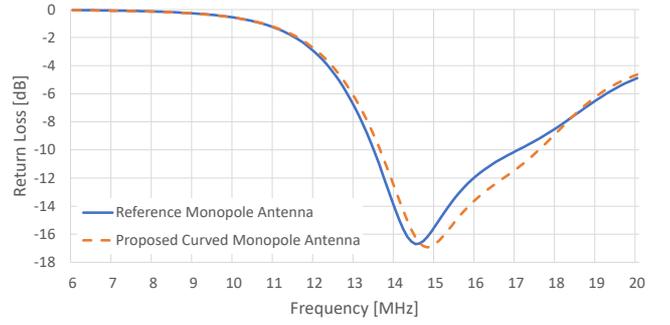

Fig. 6. Return loss vs frequency plot for the reference monopole antenna and the proposed curved monopole antenna.

broaden the operational range while maintaining compact dimensions.

Overall, these results confirm that the combination of a carefully selected straight section with a curved upper segment allows the monopole to achieve superior performance in terms of gain, bandwidth, and return loss. The proposed design achieves a more compact footprint than a conventional quarter-wavelength monopole while enhancing HF radar antenna performance, making it well-suited for practical radar deployments.

## V. ARRAY CONFIGURATION

To further evaluate the suitability of the proposed curved monopole antenna for HF skywave radar applications, the optimized single-element design was extended to a linear 12-element linear array configuration. Array operation is fundamental in HF radar systems, as it enables spatial filtering, enhanced directivity, and improved signal-to-noise ratio through beamforming and coherent processing. One primary objective of this study was to assess how the performance advantages observed at the element level can translate to array-level behavior in the elevation angle range relevant to skywave propagation.

The array consists of 12 identical curved monopole elements arranged in a linear geometry on the x-axis ($\phi = 0$) with uniform inter-element spacing of 0.45λ (where λ is the wavelength), corresponding to 9 m at the 15 MHz center frequency. This spacing was selected as a practical compromise between reducing mutual coupling and avoiding grating lobes across the operating HF band, while maintaining a physically-realizable array footprint for radar-scale installations. Each element preserves the same optimized fixed and curved geometry identified in the single-element study, ensuring that array-level performance variations arise primarily from electromagnetic interactions between elements rather than changes in individual radiator characteristics.

Particular emphasis was placed on the low-to-moderate theta angle region (θ = 0° - 45°), which is critical for skywave over-the-horizon radar operation. In this angular range, the proposed curved monopole array consistently demonstrates higher realized gain compared to a reference conventional monopole array. At an elevation angle of θ = 30°, the proposed 12-element linear array achieves a realized gain of 14.04 dBi, compared to 13.11 dBi for the reference monopole array, corresponding to 0.93 dB improvement as shown in Fig. 7 and 8. This represents approximately a 24% increase in radiated power density, which is particularly significant in HF radar systems operating at very high transmitted power levels, where an even modest gain enhancements can translate into substantial improvements in effective radiated power and system-level performance. In addition to the gain enhancement, the proposed array exhibits a lower back-lobe level, indicating improved front-to-back radiation characteristics and reduced backward radiation.

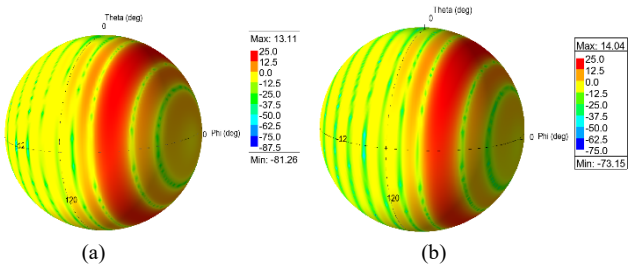

Fig. 7. 3D polar plot of the realized gain of the array of a) reference monopole antenna, and b) proposed curved monopole antenna array.

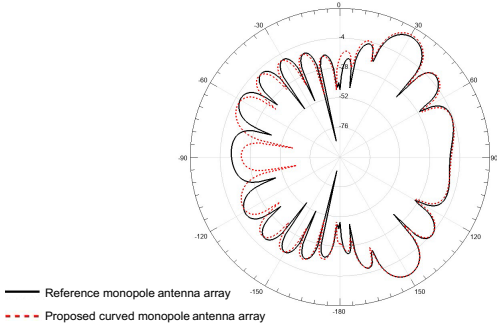

Fig. 8. Radiation pattern plot comparison in the $\varphi = 0°$ plane with the main beam steered to $\theta = 30°$.

Overall, the 12-element linear array analysis demonstrates that the proposed curved monopole antenna is not only effective as a standalone radiator but also well suited for scalable array implementations targeting skywave HF radar applications. The combination of enhanced low-elevation gain, enlarged bandwidth, and stable embedded-element performance makes the proposed design a strong candidate for next-generation HF radar systems requiring long-range coverage and directional control.

## VI. CONCLUSION

A novel curved monopole antenna for HF radar applications was presented and analyzed, with independent optimization of the straight-section length and curvature of the variable section. Parametric analysis revealed that curving only a portion of the monopole, starting from a specific height and up to an optimal curvature, can enhance radiation performance, leading to improved impedance matching, broader bandwidth, and higher realized gain. At 15 MHz, the optimized single element achieved a 18.5% gain improvement and more than 400 kHz wider bandwidth compared to a conventional quarter-wavelength monopole. Extension to a 12-element linear array with 0.45λ spacing demonstrated stable embedded-element behavior and enhanced low-to-moderate elevation gain, achieving 14.04 dBi at $\theta = 30°$ versus 13.11 dBi for the reference array, highlighting the practical advantage for skywave over-the-horizon radar systems. The proposed design thus combines compact geometry, broadband operation, and improved directional performance, making it a strong candidate for scalable HF radar arrays.

REFERENCES

[1] G. J. Frazer and C. G. Williams, "Emerging trends in radar: HF skywave radar," *IEEE Aerospace and Electronic Systems Magazine*, vol. 40, no. 6, pp. 78–82, 2025.

[2] M. Rozel, F. Yousfi, P. Brouard, L. Perus, M. Bourret and M. Menelle, "A New Generation of HF Surface Wave Radar," *2025 IEEE Radar Conference (RadarConf25)*, Krakow, Poland, 2025, pp. 165-170,

[3] J. D. Hawkins, P. V. Brennan, K. W. Nicholls, and L. B. Lok, "Design of an HF-VHF ice penetrating synthetic aperture radar," in *IEEE MTT-S International Microwave Symposium Digest*, 2022, pp. 218–221.

[4] J. D. Hawkins, L. B. Lok, P. V. Brennan and K. W. Nicholls, "HF Wire-Mesh Dipole Antennas for Broadband Ice-Penetrating Radar," *IEEE Antennas and Wireless Propagation Letters*, vol. 19, no. 12, pp. 2172-2176, Dec. 2020

[5] J. Craig, E. Gill, and R. Shahidi, "Design of high frequency surface wave radar receive monopole array elements," in *2025 IEEE International Symposium on Antenna Technology and Applied Electromagnetics*, pp. 258–261.

[6] T. Guerra-Huaranga, R. Rubio-Noriega and M. Clemente-Arenas, "Comparative Analysis of Three Types of VHF/UHF Antennas for GPR Array," in *2020 IEEE MTT-S Latin America Microwave Conference (LAMC 2020)*, Cali, Colombia, 2021, pp. 1-4.

[7] S. Yen, D. S. Filipovic and J. T. Logan, "Design of a Small Array of Electrically Small Umbrella Monopoles for HF Communications," in *2024 IEEE International Symposium on Phased Array Systems and Technology (ARRAY)*, Boston, MA, USA, 2024, pp. 1-6.

[8] S. Coutts, J. Eisenman, J. Jovin and W. Borer, "Transportable High Frequency Transmit Array with High Gain and Wide Bandwidth," *2025 IEEE International Radar Conference (RADAR)*, Atlanta, GA, USA, 2025, pp. 1-6.

[9] S. Yen, L. B. Boskovic and D. S. Filipovic, "An HF-Band Linear Retrodirective Array with Scale Model RCS Measurements," in *IEEE Transactions on Antennas and Propagation,* vol. 72, no. 7, pp. 5822-5832, July 2024.

[10] A. G. Koutinos, C. L. Zekios and S. V. Georgakopoulos, "A Monopole Cavity-Backed HF Antenna Array," in *2024 IEEE International Symposium on Phased Array Systems and Technology (ARRAY)*, Boston, MA, USA, 2024, pp. 1-3.